\documentclass[10pt,
							 twocolumn,
							 superscriptaddress,
							 english,
							 prl,
							 showpacs,
							 floatfix,
							 aps
							]{revtex4-1}
\usepackage[utf8]{inputenc}
\usepackage{amsmath}
\usepackage{amssymb}
\usepackage{graphicx}
\usepackage{xspace}
\usepackage{braket}
\usepackage{color}
\makeatletter

\begin{document}

\title{Fracton Critical Point in Higher-Order Topological Phase Transition}

\author{Yizhi You}
	\affiliation{Princeton Center for Theoretical Science, Princeton University, NJ, 08544, USA}
\author{Julian Bibo}
	\affiliation{Department of Physics, Technical University of Munich, 85748 Garching, Germany}
	\affiliation{Munich Center for Quantum Science and Technology (MQCST), D-80799 Munich, Germany }

\author{Frank Pollmann}
	\affiliation{Department of Physics, Technical University of Munich, 85748 Garching, Germany}
	\affiliation{Munich Center for Quantum Science and Technology (MQCST), D-80799 Munich, Germany }
	\author{Taylor L. Hughes}
\affiliation{Department of Physics and Institute for Condensed Matter Theory, University of Illinois at Urbana-Champaign, Illinois 61801, USA}

\date{\today}

\begin{abstract}
The theory of quantum phase transitions separating different phases with distinct symmetry patterns at zero temperature is one of the foundations of modern quantum many-body physics. In this paper we demonstrate that the existence of a 2D topological phase transition between a higher-order topological insulator (HOTI) and a trivial Mott insulator with the same symmetry eludes this paradigm. We present a theory of this quantum critical point (QCP) driven by the fluctuations and percolation of the domain walls between a HOTI and a trivial Mott insulator region. Due to the spinon zero modes that decorate the rough corners of the domain walls, the fluctuations of the phase boundaries trigger a spinon-dipole hopping term with fracton dynamics. Hence we find the QCP is characterized by a critical dipole liquid theory with subsystem $U(1)$ symmetry and the breakdown of the area law entanglement entropy which exhibits a logarithmic enhancement: $L \ln(L)$. Using the density matrix renormalization group (DMRG) method, we analyze the dipole stiffness together with structure factor at the QCP which provide strong evidence of a critical dipole liquid with a Bose surface. These numerical signatures further support the fracton dynamics of the QCP, and suggest a new paradigm for 2D quantum criticality proximate to a topological phase.
\end{abstract}

\maketitle

{\em \textbf{Introduction~---}} After a decade of intense effort focused on topological insulator materials, a new class of symmetry protected topological insulators, dubbed Higher-Order Topological Insulators~(HOTI) has been discovered~\cite{benalcazar2017quantized,schindler2017higher,langbehn2017reflection,song2017d}. HOTIs admit gapped surfaces separated by gapless corners/hinges where the surfaces intersect, and exemplify a rich bulk-boundary correspondence. Aside from HOTIs generated by topological band structures, recent research suggests strongly interacting bosonic systems can potentially host a HOTI having robust bosonic corner zero-modes~\cite{you2018higherorder,dubinkin2018higher}. Now the classification and characterization of interacting HOTIs has been widely-explored in terms of mathematical invariants, topological response, and field theory approaches~\cite{rasmussen2018classification,you2019multipolar}.
However, the existence and character of a quantum phase transition between an interacting HOTI phase and a trivial Mott insulator phase is still nebulous. In particular, it is noteworthy to explore whether the critical region inherits the topological properties of the HOTI, and how such a phase transition is influenced by the topological structure, and the entanglement pattern, of the adjacent HOTI phase.

In this work, we address these questions and propose a new type of quantum critical point that connects a 2D HOTI phase~\cite{you2019multipolar,bibo2019fractional} and a trivial Mott insulator phase. The traditional guiding principle behind the modern theory of critical phenomena, known as the Ginzburg-Landau-Wilson (GLW) paradigm, is the identification of an `order parameter fluctuation’ that encapsulates the differences in symmetry between the two phases proximate to the critical point. However, the quantum critical point (QCP) we present in this paper eludes this paradigm as both the HOTI phase, and the trivial Mott phase, have the same symmetries, and are distinguished through only their different topological character.

Remarkably, we find that the phase transition we propose inherits topological features from the HOTI phase. The QCP can be understood as the bulk percolation~\cite{chen2013critical} of domain walls that act as phase boundaries between regions containing HOTI or trivial Mott insulator. The corners and rough patches of the 1D domain walls can be treated as the corners of the HOTI phase, and are thus each decorated with a robust spinon zero mode. At the QCP, the proliferation of domain walls triggers the fluctuations of the corner spinon zero modes, and precipitates fracton dynamics of the spinons that are constrained by subsystem $U(1)$ symmetry. Hence, we find that this critical point contains quasiparticles with fracton behavior and sub-dimensional kinetics where the spinon-dipoles only move transverse to their dipole moment~\cite{Vijay2015-jj,Vijay2016-dr,Chamon2005-fc,Haah2011-ny,yoshida2013exotic,pretko2020fracton}.  This new type of quantum criticality leads us to propose that the phase transition is characterized by a critical dipole liquid~\cite{you2019emergent,xu2007bond,paramekanti2002ring,tay2010possible,mishmash2011bose}. This critical theory has several key features including: (i) a Bose surface~\cite{sachdev2002scratching} having zero energy states that form closed nodal-lines along the $k_x$ and $k_y$ axes, and (ii) a breakdown in the area law of the entanglement entropy, which is replaced by a scaling with a logarithmic enhancement $L \ln(L)$ at the critical point instead. 
To our knowledge, this is the first topological phase transition in 2D having entanglement entropy that exceeds the area law~\cite{lai2013violation}.

\begin{figure}[h]
\includegraphics[width=0.48\textwidth]{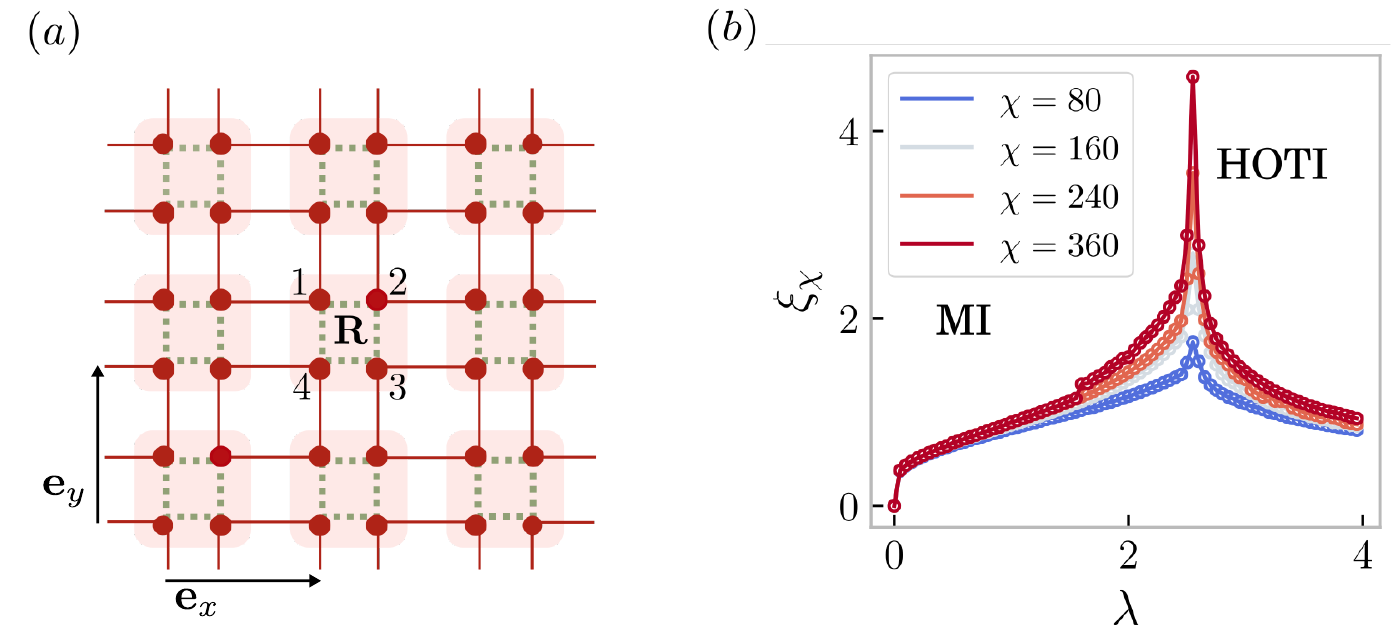}
\caption{\textbf{Lattice model and quantum critical point (QCP).} Panel $(a)$ shows the lattice model. Each unit cell consists of four spin $S=1/2$ (red dots). The ring-exchange terms (red squares) couple four neighboring unit cells, while sites in the unit cell are isotropically coupled via an $XY$ interaction (green dotted lines). Panel (b) shows the correlation length $\xi_\chi$ as function of the tuning parameter $\lambda$ and the bond dimension $\chi$ for an infinite cylinder along $x$, $L_x=\infty$, and periodic boundary conditions along $y$ with a circumference $L_y =6$. A second order quantum phase transition between a Mott-Insulator (MI) and higher order topological insulator (HOTI) occurs at $\lambda_c \approx 2.54$.} 
\label{phase}
\end{figure}

{\em \textbf{HOTI transition}~---} To frame our discussion we consider the following model on a 2D square lattice with four spin-1/2 degrees of freedom per unit cell:  
\begin{align}  
&H=H_{XY}-\lambda H_{\text{ring exchange}}\nonumber\\
&= \sum_{{\bf{R}}}\left(S^{+}_{{\bf{R}},1} S^{-}_{{\bf{R}},2}+S^{+}_{{\bf{R}},2} S^{-}_{{\bf{R}},3}\right.
\left.+S^{+}_{{\bf{R}},3} S^{-}_{{\bf{R}},4}+S^{+}_{{\bf{R}},4} S^{-}_{{\bf{R}},1}\right)\nonumber\\
&-\lambda\sum_{{\bf{R}}}\left( S^{+}_{{\bf{R}},2} S^{-}_{{\bf{R}}+\mathbf{e}_x,1} S^{+}_{{\bf{R}}+\mathbf{e}_x+\mathbf{e}_y,4} S^{-}_{{\bf{R}}+\mathbf{e}_y,3}\right)+h.c.
\label{hhh}
\end{align}
Hamiltonian~\eqref{hhh} contains an inter-cell ring-exchange interaction between the four spins located at the four corners of each red plaquette, and an XY spin interaction within the unit cell as shown in Fig.~\ref{phase}(a). In passing we mention that models having ring-exchange terms of this type can be realized in cold-atom settings, which hence is a natural arena for experimental investigation of our subsequent predictions~\cite{paredes2008,dai2017}. 
The magnon creation/annihilation operators $S^{\pm}=\sigma^x\pm i\sigma^y$ can be mapped to a hardcore boson description using $b^{\dagger}(b)=S^+(S^-), S^z=n_b-1/2$, where $n_b=b^{\dagger}b^{\phantom{\dagger}}$; we use both languages interchangeably where convenient.
Eq.~\eqref{hhh} exhibits time-reversal $\mathcal{T}=\prod_{\bf{R}}\prod^{4}_{m=1}i\sigma^y_{\mathbf{R},m}\mathcal{K}$ symmetry and a subsystem U(1) symmetry that conserves the sum of $S^z$ inside the unit cell $\mathbf{R}$, i.e. $S^z(\mathbf{R})=\sum^4_{m=1}S^z_{\mathbf{R},m},$ for every row (R) and column (C),
\begin{align} 
& U^{\mathrm{sub}}_{R(C)}(1): \prod_{\mathbf{R} \in R(C)} e^ {i \theta  S^z(\mathbf{R})}.
\end{align} 
The subsystem U(1) symmetry restricts the mobility of the magnon, hence the leading order dynamics are attributed to  pairs of dipoles, each composed of a particle-hole pair on a lattice link, that hop along the direction transverse to their dipole moment. 

 Before turning to the detailed study of the phase diagram for the model Eq.~\eqref{hhh}, let us discuss two exactly solvable limits. As $\lambda$ is tuned, the system effectively has competing orders dominated by either the intra-cell XY interaction, or the inter-cell plaquette ring exchange interaction. In the limit $\lambda \sim 0$, the intra-cell term plays the key role and generates an entangled cluster within each unit cell. The resulting ground state, which is simply a tensor product of the ground state of each unit cell, is a featureless Mott insulator with a magnon gap for both the bulk and boundary.
In the opposite limit $\lambda \sim \infty$, it was shown in Ref.~\cite{you2019multipolar} that
the plaquette term projects the four interacting spins into a unique maximally entangled state
$|\downarrow_2 \uparrow_1 \downarrow_4 \uparrow_3 \rangle+|\uparrow_2 \downarrow_1 \uparrow_4 \downarrow_3\rangle$
where we omitted unit cell labels.
The corresponding ground state for the entire system is thus a product of entangled plaquettes, and has a finite magnon gap in the bulk. In the presence of a smooth boundary, each edge unit cell naively has two free spin-1/2 modes, but these can be coupled and gapped to form a 
singlet state via the onsite XY coupling in Eq.~\eqref{hhh} (even for infinitesimal $\lambda$). For rough edges and/or corners, there is an additional spin-1/2 zero mode per corner whose two-fold degeneracy is protected by $\mathcal{T}$ and subsystem U(1) symmetry. Based on this observation, this gapped ground state is a higher-order topological insulator with robust corner modes protected by $\mathcal{T}$ and subsystem U(1) symmetry. This subsystem symmetric HOTI phase was studied in Ref.~\cite{you2019multipolar}, and was shown to exhibit a quantized quadrupole moment density of $Q_{xy}=1/2$. 

We obtained the phase diagram, Fig.~\ref{phase}(b) using the DMRG method~\cite{White1992, Mcculloch2008, Hauschild2018} on an infinitely long cylinder. We find that the aforementioned exactly solvable limits extend into two gapped phases (a trivial Mott insulator and a HOTI respectively). Importantly, our numerics indicate that the two gapped phases are connected by a second-order phase transition at $\lambda_c\approx2.54$. It is noteworthy to emphasize that the HOTI phase and the trivial Mott phase display the same symmetries, but harbor distinct topological features, and thus cannot be differentiated via any local observable. This further implies that the QCP between the phases cannot be accessed by a GLW type phase transition theory based on a fluctuating order parameter.

\begin{figure}[h]
\includegraphics[width=0.48\textwidth]{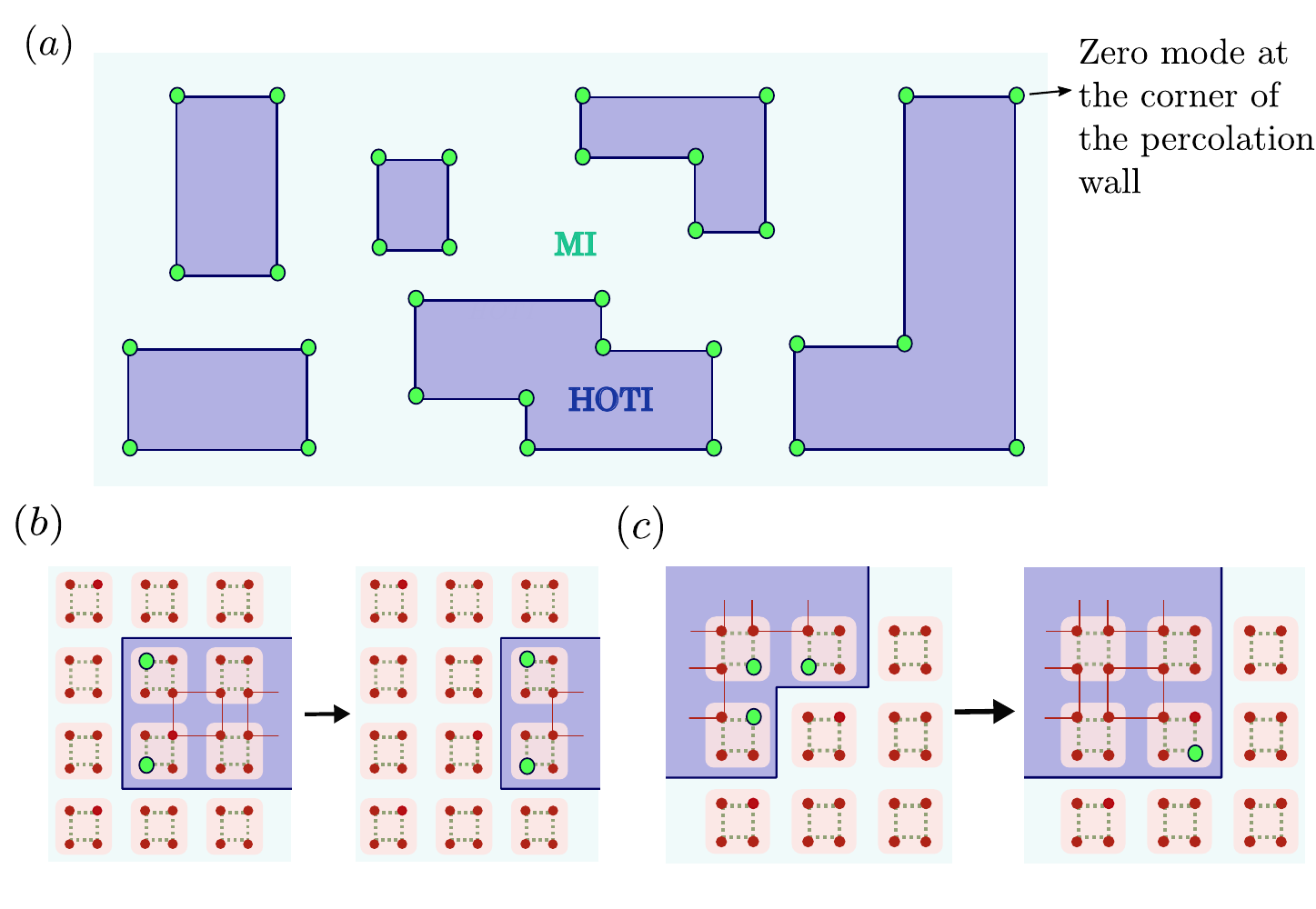}
\caption{(a) At the critical point, the HOTI~(blue shaded area) and trivial Mott insulator (green shaded region) coexist and their phase boundary~(domain wall) contains spinon zero modes (green dot) at the corners. (b)-(c) The spatial fluctuation of phase boundaries (domain walls) induces motion for the spinon-pairs at neighboring corners.}
\label{peroco}
\end{figure}

{\em \textbf{Critical theory}~---}
We now provide both an analytic argument and numerical evidence to demonstrate that the quantum critical point separating the HOTI phase and the trivial Mott phase displays gapless fracton quasi-particles akin to a critical dipole liquid. When the interaction strength $\lambda$ is comparable to the intra-cell tunneling strength, the plaquette entangled patterns and on-site entangled patterns compete and coexist in the bulk. The coexistence and spatial phase separation can be viewed from a percolation picture illustrated in Fig.~\ref{peroco}. In the quantum critical region adjacent to the trivial Mott phase, the plaquette ring-exchange term triggers some regions containing plaquette entangled states that exhibit the HOTI ground state pattern. Domain walls that form between the two phases  can be viewed as the boundary between the HOTI and trivial phases, and hence they harbor a spinon zero mode at each corner (and each `rough' patch on the edges). In the quantum critical region, strong fluctuations between different phase separation patterns are induced, and the domain walls tend to proliferate. These spatial fluctuations concurrently trigger the dynamics of the spinon zero modes on the corners of the domain walls, similar to the percolation of domain wall defects in 1D~\cite{chen2013critical}.

More precisely, the spinon dynamics originates from the resonance between distinct percolation patterns. %
In Figs.~\ref{peroco}(b) and (c) we display two typical phase boundary deformations. By shrinking a stripe domain along the $x$-direction, the two spinon zero modes forming an effective $y$-oriented dipole can hop along the $x$-direction, and vice-versa. This motion can be described by a ring exchange term for the spinons:
\begin{align} 
& \sum^{2}_{i=1} z^{\dagger}_i(\mathbf{R})z_i(\mathbf{R}+\mathbf{e}_x)z^{\dagger}_i(\mathbf{R}+\mathbf{e}_x+\mathbf{e}_y)z_i(\mathbf{R}+\mathbf{e}_y),
\label{spinon}
\end{align} where $(z^{\dagger}_{1},z^{\dagger}_{2})$ is the $CP^1$ representation of the spinon with $S^z_{i}=\tfrac{1}{2}z^{\dagger}_{i}\sigma^z_{ij} z_{j}$. 
Another typical domain wall deformation occurs when a corner is shrunk by removing a corner plaquette from the stripe. This effectively removes the free spinon from the corner, but it also creates three other spinons at the newly created corners. This process is also represented by the ring-exchange term in Eq.~\ref{spinon}. 
Hence, the spatial fluctuations of the percolating domain walls generate a ring-exchange type term for spinons that effectively corresponds to the motion of dipoles transverse to their dipole moment. An important characteristic of this transition is that the percolation at the QCP does not trigger the hopping of a single spinon, stemming from the fact that the spinon modes on the phase boundaries are localized at the corners. 

Based on these key observations, we propose that the low energy effective description of the QCP is a critical dipole liquid theory composed of spinons :
\begin{align} 
&\mathcal{L}=\sum_{\gamma=1,2} (\partial_t \theta_\gamma+a_0)^2-K (\partial_x \partial_y \theta_\gamma+a_{xy})^2,\nonumber\\
&a_{xy}\rightarrow a_{xy}+\partial_x\partial_y \alpha, ~a_{0}\rightarrow a_{t}+\partial_t\alpha,~~
z^{\dagger}_i={n}_i e^{i \theta_i}.
\label{ac}
\end{align}
Here we have introduced a number-phase representation for the $CP^1$ spinons, and $a_0,a_{xy}$ are components of the emergent gauge field that couple to the gauge charge of the spinon; their gauge transformations are also listed above. We note that we are ignoring the compactification of the boson fields $\theta_i$, and have expanded to quadratic order in them~\cite{you2019emergent}. The legitimacy  of this approximation, which is tied to the irrelevance of instanton tunneling events, is discussed in detail in the Appendix.  To further analyze this theory, we can decompose the two $CP^1$ phase fields as $\theta_{\pm}=\theta_1 \pm \theta_2$ to find:
\begin{align} 
\mathcal{L}=\frac{1}{2}\sum_{\gamma=\pm}(\partial_t \theta_\gamma)^2-\frac{K}{2}(\partial_x \partial_y \theta_+ +a_{xy})^2-\frac{K}{2}(\partial_x \partial_y \theta_-)^2.
\label{2dspin23}
\end{align}
We find that the field $\theta_+$ is the only bosonic spinon mode that couples to the emergent gauge field $a_{xy}$. Hence this mode is gapped out due to a Higgs-like mechanism for the gauge field, or equivalently, by the onsite XY interaction (see Appendix for details). The field $\theta_-$ denotes the gapless magnon mode: $S^{\pm}=e^{i\pm \theta_-}$ that carries the $S_z$ quantum number. This field contributes to the low energy dynamical phenomena at quantum criticality, and henceforth we only focus on the $\theta_-$ branch.
 
 Due to subsystem $S_z$ conservation, the action in Eq.~\eqref{2dspin23} is invariant under a special U(1) transformation $
\theta_- \rightarrow \theta_- +f(x)+g(y).$
Consequently, the single magnon hopping term $(\partial_i \theta_-)^2$ is forbidden as it breaks this subsystem U(1) symmetry explicitly. Instead, the leading order dynamics originates from dipole moments oriented along the $i$-th direction, i.e., ($ \partial_i \theta_{-}$) that are constrained to move along the transverse $j$-th direction.   This is captured in our theory by a special, higher order kinetic term $(\partial_j\partial_i\theta_{-})^2$ yielding a dispersion $\omega \sim k_x k_y.$  This is remarkable because we find that the QCP exhibits  characteristic fractonic features arising from a percolation process where the corner-localized low-energy modes cannot fluctuate independently without creating additional corners. Finally, we note that this theory can also be used to describe the gapped phases proximate to the QCP (see Appendix for details).

  An alternative way to illustrate the fracton dynamics at the critical point is to explore the magnon dynamics directly from the microscopic Hamiltonian in Eq.~\eqref{hhh}. When the interaction strength $\lambda$ grows, the inter-cell coupling term triggers the magnons to strongly fluctuate between unit cells. However, as the inter-cell coupling contains only a ring exchange term, it does not support single charge transport between cells, and hence the leading order dynamics is governed by a pair of magnon-dipoles moving between cells. This sub-dimensional mobility prevents spontaneous symmetry breaking of the subsystem $U(1)$ symmetry for $S_z$ due to the Mermin-Wagner theorem~\cite{Batista2005-yr,tay2010possible,paramekanti2002ring,seiberg2020exotic}. Thus, we would not expect a superfluid instability at the critical point. To corroborate this we find that the two point correlator $\langle S^+(\mathbf{R}) S^-(\mathbf{R})\rangle$ is short ranged at the QCP. In contrast, since we expect that the critical region manifests a critical dipole liquid, we find that the four-point correlation functions between two spin-dipoles living on the same transverse stripe, exhibit~\cite{you2019emergent}:
\begin{equation}
\langle S^{+}(\mathbf{R}) S^{-}(\mathbf{R}+\mathbf{e}_{y}) S^{-}(\mathbf{R}+x) S^{+}(\mathbf{R}+x+\mathbf{e}_y) \rangle 
=\frac{1}{(x)^{1/(K\pi^2)}}, 
\end{equation}
and analogously for a stripe along $y$.
These correlation functions exhibit algebraic decay and quasi-long range order if and only if the two dipoles are living on the same stripe. This again supports the fact that we should interpret the critical point as a dipole liquid having  constrained subsystem dynamics.

\begin{figure}[h]
\includegraphics[width=0.48\textwidth]{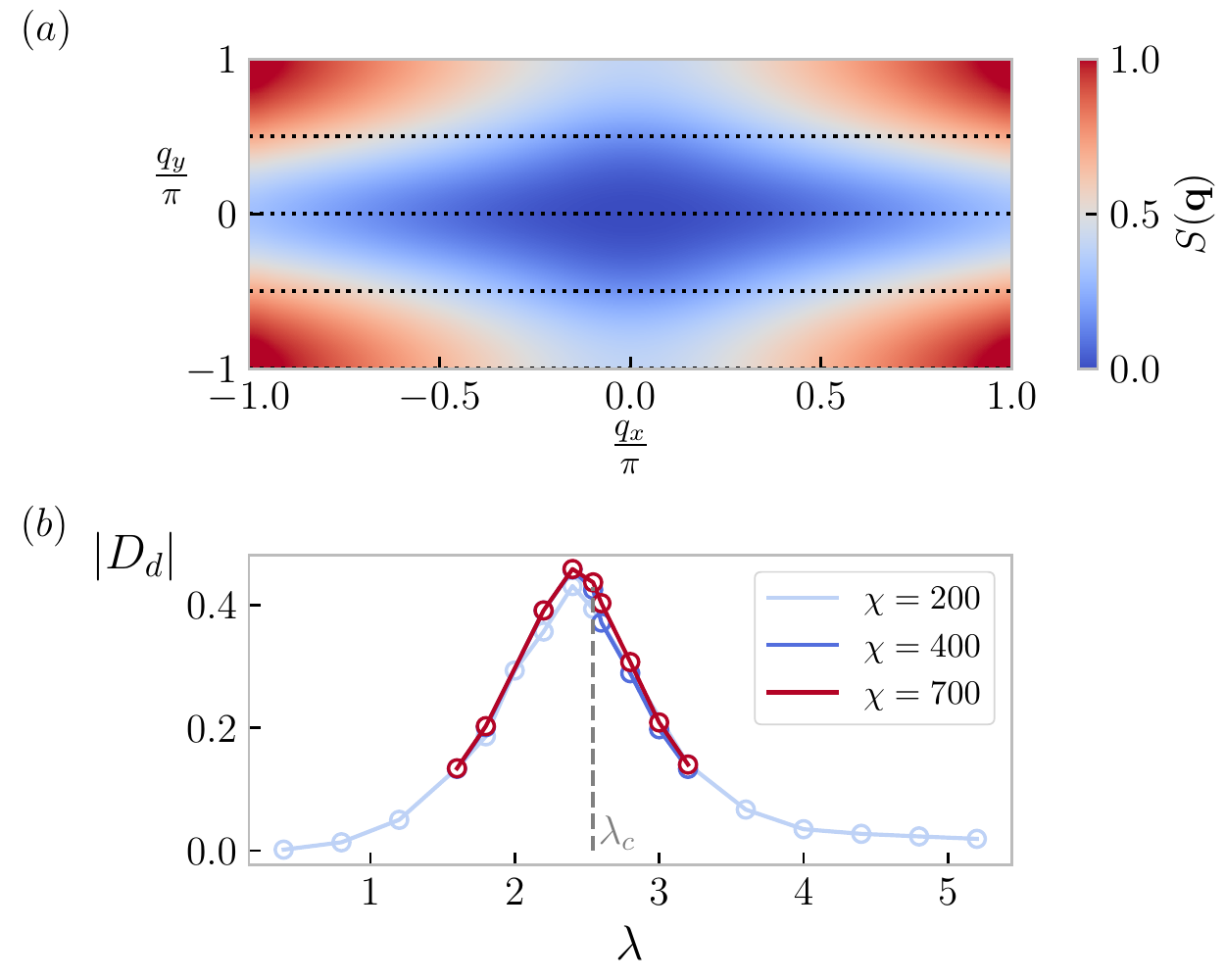}
\caption{\textbf{Static structure factor and dipole stiffness.} $(a)$ The static structure factor $S(\mathbf{q})$ calculated for a stripe with $L_x = 100$ and $L_y = 8$ sites along the $x$ and $y$ direction, respectively. Hence, the data are obtained for momenta $q_y =0,\pm \pi,\pm \pi/2$ (black, dotted lines). $(b)$ The absolute value of the dipole stiffness $|D_d|$, calculated for an infinite cylinder with $L_y = 6$. The dipole stiffness is obtained by taking the second, symmetric derivative with step size $\delta\Phi = 0.05$.}
\label{data}
\end{figure}

{\em \textbf{Properties of QCP}~---}
The critical theory in Eq.~\eqref{2dspin23} produces a quadratic dispersion $\omega \sim k_x k_y$, which implies a Bose surface with characteristic lines of zero energy modes on both the $k_x$ and $k_y$ axes.  For each fixed momentum slice $k_i \neq 0$, the low energy dispersion is akin to a conventional 1D relativistic boson moving along the transverse direction. Such `quasi-1D' motion is a consequence of the sub-dimensional nature of the critical dipole liquid, i.e., the fact that  an $x(y)$-oriented dipole is only mobile along the transverse $y(x)$-oriented stripes. 
To confirm the existence of the Bose surface we numerically evaluate the structure factor:
\begin{align}
S(\mathbf{q})=&\frac{1}{N}\left[\prod^2_{m=1}\sum_{\mathbf{R}_m,\mathbf{b}_m}e^{i\mathbf{q}\cdot\left[(-1)^m\mathbf{\Lambda}_m\right]}\right]\braket{{S}^{z}_{\mathbf{\Lambda}_1}{S}^{z}_{\mathbf{\Lambda}_2}},
\end{align}
where $N=L_y L_x/4$ is the total number of unit cells, $\mathbf{\Lambda}_m = \mathbf{R}_m+\mathbf{b}_m$ specifies one the four spins $S^z_{\mathbf{R}_m,m}$ inside the unit cell $\mathbf{R}_m$ with basis vectors $\mathbf{b}_m\in\frac{1}{2}\{\mathbf{0},\mathbf{e}_x,\mathbf{e}_y,\mathbf{e}_x+\mathbf{e}_y\}$. 
In Fig.~\ref{data}(a) we show that the numerically obtained structure factor exhibits clear zero-energy lines along the $k_x,k_y$ axes. Furthermore, each $k_i \neq 0$ exhibits dispersion like a 1D relativistic boson along the transverse direction. We note that since there are nodal lines at $k_x, k_y =0,$ there is a sub-extensive number of quasi-1D modes, and the specific heat at low temperature will scale as $C_v \sim T \ln(1/T),$ which is similar to marginal Fermi liquid theory in 2D~\cite{you2019emergent,paramekanti2002ring,xu2007bond}.

As a further confirmation of the dipole liquid critical point,  we can also characterize the dipole liquid by its transport properties since our model has dipole conservation. In the language of Ref.~\cite{dubinkin2019theory}, we expect the critical dipole liquid to act as a `dipole metal' and exhibit a non-vanishing dipole conductivity in the presence of a uniform rank-2 electric field, e.g., $E_{xy}.$ In analogy to Kohn's seminal work on defining conductors and insulators~\cite{kohn1964theory}, Ref.~\cite{dubinkin2019theory} proposed a criterion to establish the existence of a dipole metal. This criterion is based on the dipole stiffness, defined as
\begin{equation}
    D_d=-\frac{\pi}{V}\frac{\partial^2 E_0}{\partial \mathfrak{q}^2},
\end{equation} where $E_0$ the ground-state energy, and $\mathfrak{q}$ represents a constant shift of the rank-2 gauge potential $A_{xy}=\mathfrak{q}.$ The key criterion is that if $D_d$ is non-vanishing in the thermodynamic limit, then the system is a dipole metal. Our model naturally couples to $A_{xy}$ through a Peierls factor on ring-exchange terms
$
 S^{+}_{2} S^{-}_{1}S^{+}_{4} S^{-}_{3}\to e^{iA_{xy}} S^{+}_{2} S^{-}_{1} S^{+}_{4} S^{-}_{3}
$~\cite{you2019multipolar},
where we again omitted unit cell labels. Hence, we can use the sensitivity of the QCP to rank-2 boundary condition twists to determine if it is a dipole metal as we expect. Numerically, the dipole stiffness can be obtained by a two-step process: (i) we multiply each ring-exchange term by a phase factor $e^{4\pi i\frac{\Phi}{L_y}}$~\footnote{Note the number of unit cells along $y$ is equal to $L_y/2$}, (ii) we take the second, symmetric derivative of the flux-dependent ground state energy $E_0(\Phi)$ to obtain $D_d$, which is shown in Fig.~\ref{data}(b). Our numerical results clearly support our theory since the system shows a non-vanishing $D_d$ only in the neighborhood of the critical point. We note that although the parameter value of the QCP obtained via the correlation length and the dipole stiffness differ slightly from each other at $\lambda_c=2.5$, both suggest a single QCP. The deviation is a numerical artifact since DMRG at a QCP in 2D suffers not only from the effective long range couplings, but also from the enormous amount of entanglement at this particular, quantum critical point.

Indeed, another noteworthy feature for the critical dipole liquid is an unusually large entanglement entropy scaling that exceeds the area law.
In conventional 2D quantum critical points, despite the divergent correlation length, the entanglement entropy of the ground state carried by a sub-region of the many-body system is still local in the sense that it scales with $L,$ the perimeter of the subsystem boundary. However, in the critical theory we present here, the entanglement entropy should scale with the sub-region size as $L \ln(L)$~\cite{boseentanglement,swingle}.  
Such a violation of the area law can be substantiated by dividing the Bose surface into small patches over which the surface looks approximately flat. Each patch can be regarded as a 1D relativistic boson whose entanglement entropy scales as $\ln(L)$. Summing over the contributions from all patches, the total entanglement entropy should scale as $L \ln(L)$. To our knowledge, this is the first observation of a 2D quantum critical point whose entanglement pushes beyond the area law. This implies that the critical dipole liquid has long range mutual information shared by two regions far apart.

  {\em \textbf{Discussion}~---} 
  We provided a framework to describe a novel type of 2D quantum criticality with logarithmic entanglement scaling and emergent fracton dynamics in the absence of Lorentz invariance. Our description of the HOTI percolation transition also suggests new insights for various topological phase transitions and critical points beyond the GLW paradigm, and is certain to have rich theoretical and experimental implications.
  While our numerical DMRG simulations are implemented on cylinders with finite circumference, the present data still provide strong indications for a new type of 2D QCP. In particular, the critical dipole liquid theory we propose is  sub-dimensional since algebraically decaying correlations and a diverging correlation length exist only among dipoles on the same stripe. This quasi-1D dynamics enables us to access the 2D critical point from finite stripe simulations. We also expect that the model we have studied may be realizable in near-term cold-atom experiments.

{\em \textbf{Acknowledgments}~---}This work is initiated at KITP and YY, TLH, FP are supported in part by the National Science Foundation under Grant No. NSF PHY- 1748958(KITP) during the Topological Quantum Matter program. FP acknowledges the support of the DFG Research Unit FOR 1807 through grants no. PO 1370/2-1, TRR80, and the Deutsche Forschungsgemeinschaft (DFG, German Research Foundation) under Germany's Excellence Strategy EXC-2111-390814868. TLH thanks the US National Science Foundation (NSF) MRSEC program under NSF Award Number DMR-1720633 (SuperSEED) and the award DMR 1351895-CAR for support.

\clearpage

\setcounter{figure}{0}
\setcounter{section}{0}
\setcounter{equation}{0}
\renewcommand{\theequation}{S\arabic{equation}}
\renewcommand{\thefigure}{S\arabic{figure}}

\onecolumngrid

\newcommand{\vsigma}{\mbox{\boldmath $\sigma$}}

\section*{\Large{Supplemental Material}}

\vspace{0.7cm}

\section{Spinon at the corner}
\begin{figure}[h]
\includegraphics[width=0.3\textwidth]{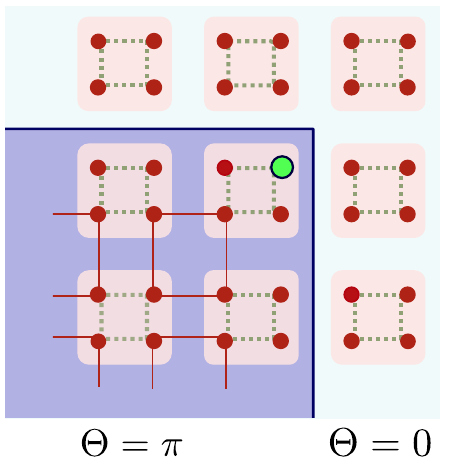}
\caption{The HOTI (grey) region and the trivial Mott (white) region can be labeled as a $Z_2$ variable $\Theta=\pi,0$ corresponds to the quadrupole moment. At the corner $e^{i (\Delta_x \Delta_y \Theta)}=-1$ imples there being a spinon charge. The smooth boundary contains two spinon which would be paired into a singlet. }
\label{cor}
\end{figure}

The corner (or rough domain wall) between the HOTI region and trivial Mott region contains a free spin-1/2 degree of freedom. If we define the HOTI and trivial region in terms of a $Z_2$ variable $\Theta=0,\pi$ corresponding to the quadrupole moment~\cite{you2019multipolar}, the corner modes implies,
\begin{align}
&e^{i (\Delta_x \Delta_y \Theta)} = e^{i\pi q_s}
\end{align}
Each corner contains an unpaired spinon $q_s=1$ as Fig~\ref{cor}. 

Here we use the $CP^1$ representation of the spinon,
\begin{align}
& S^{a}=z_i^{\dagger} \sigma^{a}_{ij} z_j,~~q_s=z_1^{\dagger} z_1+z_2^{\dagger}  z_2
\end{align}
For the smooth boundary with two spinons per unit cell, any infinitesimal intra-site XY interaction would pair them into a singlet so the spinon charge on the smooth boundary is zero. 
As any even number of unpaired spinon per unit cell can be paired into a singlet, the spinon here is only well defined modulo 2. This is also apparent from the weak intra-site XY interaction which has a tendency to pair two spins as $z^{\dagger}_1(R)z^{\dagger}_2(R+b_i)-z^{\dagger}_2(R)z^{\dagger}_1(R+b_i)$ and hence Higgs the gauge charge of the spinon from U(1) to $Z_2$. Subsequently, the $\theta^+$ mode in Eq.~\ref{ac} at the critical point is gapped and would not contribute to the low energy physics. (Such Higgsing does not affect the Magnon in the meantime as the U(1) symmetry with respect to $S_z$ charge is not broken.)
This also implies the emergent gauge field $a_{xy}$ carried by the spinon is Higgsed into a discrete $Z_2$ gauge field $a_{xy}=0,\pi$. As the higher-rank gauge field does not render any gauge fluctuation, such Higgsing would not change the critical behavior.

\section{Compactification and instanton event at QCP}
In this appendix, we address the legitimacy of the spin wave approximation in Eq.~\ref{ac}. The $CP^1$ boson at the QCP is compact and the Gaussian expansion of the kinetic theory in Eq.~\ref{ac} is only valid provided all instanton proliferation inducing the $2\pi$ tunneling of the phase field $\theta_i$ is irrelevant.

\begin{align} 
\mathcal{L}=\frac{K}{2}\sum_{a=+,-}(\partial_t \theta_a)^2-\frac{K}{2}(\partial_x \partial_y \theta_+ +a_{xy})^2-\frac{K}{2}(\partial_x \partial_y \theta_-)^2
\end{align}
Such a quadratic Lagrangian, in which all terms involve derivatives of the fields, describes a scale invariant phase at long length scales. In a sense it can be viewed as a ``fixed point" Lagrangian. The $\theta_+$ mode and the gauge field $a_{xy}$ gap out each other through an analog of the Higgs mechanism. Subsequently, only the $\theta_-$ branch is physical and that is the degree of freedom we will consider hereafter.

 Following the duality and bosonization argument introduced in Ref.~\cite{xu2007bond,paramekanti2002ring}, we can map the theory to its dual representation.  As the $\theta_-$ fields are compact with the identification $\theta_-=\theta_-+2\pi Z $, particular types of topological defects will be allowed, which can be most conveniently addressed by passing to a dual representation $N,\phi$ defined on the plaquette centers\footnote{$\hat{n}_-=s_z+1/2$ is the conjugate variable of $\theta_-$.}.

 \begin{figure}[h]
\includegraphics[width=0.3\textwidth]{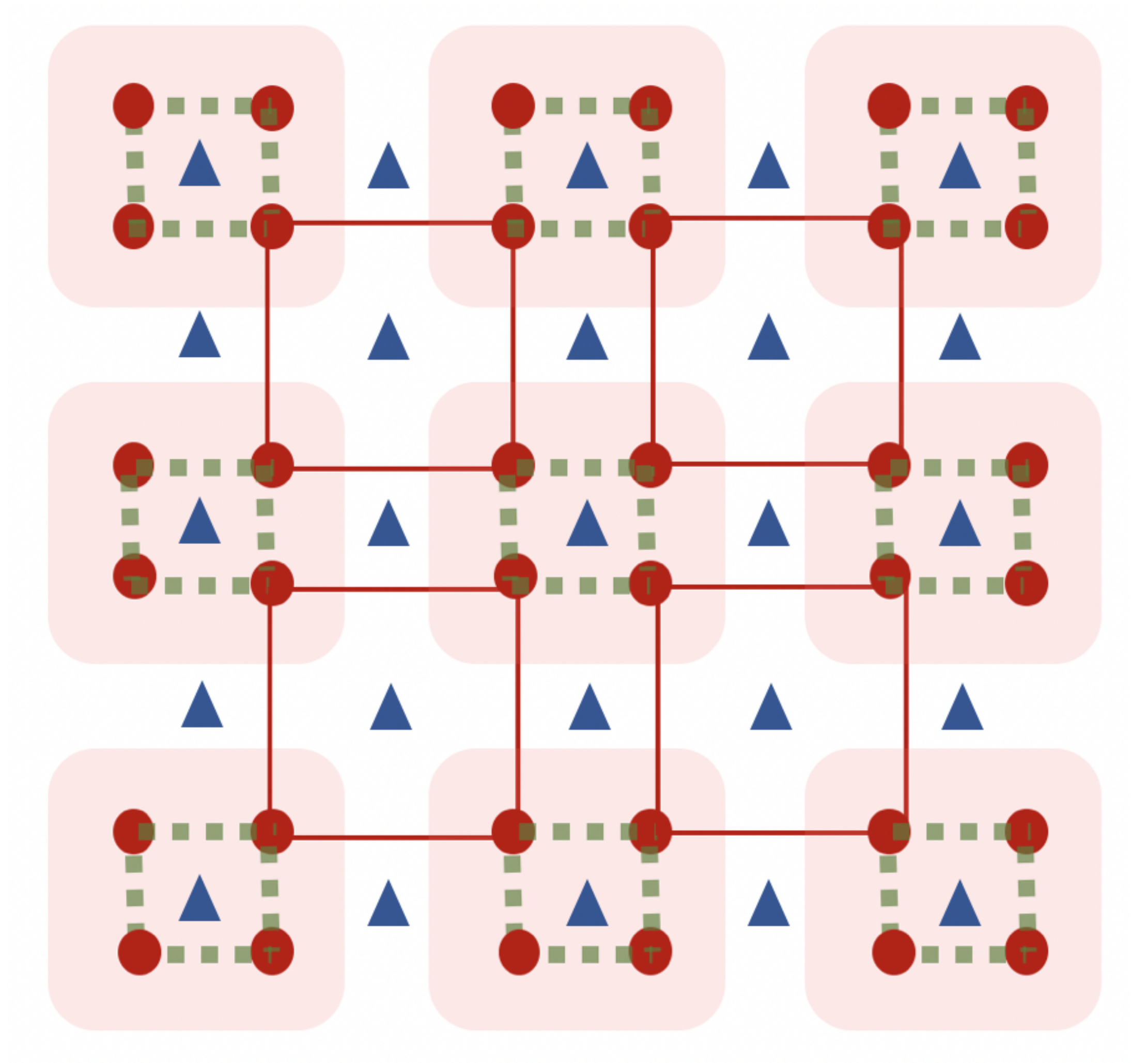}
\caption{We enlarge each unit cell into a plaquette with 4 spins spatially separated apart so the original model can be viewed as the spin-1/2 model on a super-lattice. The blue triangles are the dual variables $\phi, N$ living on the dual lattice.}
\label{app}
\end{figure}

 For simplicity, we enlarge each unit cell into a plaquette with 4 spins spatially separated apart as Fig.~\ref{app} so the original model can be viewed as the spin-1/2 model on a super-lattice.
\begin{align} 
\hat{n}_--1/2=\partial_i\partial_j \phi_-,~\hat{N}_-=\partial_i\partial_j \theta_-
\label{dual}
\end{align}
$\hat{N}$ and $\phi$ are a pair of conjugate variables with $\phi$ being discrete-valued and $\hat{N}$ being compact with $N \in [0,2\pi]$. 

The Gaussian part of the dual action is,
\begin{align} 
\mathcal{L}=\frac{1}{2K}(\partial_t \phi_-)^2-\frac{1}{2K}(\partial_x \partial_y\phi_-)^2
\label{2dspindual}
\end{align}
The $K$ can be view as the Luttinger liquid parameter whose value depends on the magnon interaction.
In this dual picture, due to the discreteness of $\phi_-$, one can also add vertex operators such as $\cos(4 \pi \partial_i \phi_-)$ at the QCP. We now demonstrate that as long as $K$ pass a critical value, the instanton operator is irrelevant so the spin wave approximation is valid.
In the original representation, the correlation between two charges $\langle \cos(\theta_-(r)) \cos(\theta_-(0))\rangle$ vanishes at long-wave length due to the subsystem $U(1)$ symmetry.  The leading order non-vanishing correlation functions are between two dipole operators living on bonds,
\begin{align} 
&\langle \cos(\partial_x\theta_-)(0,0,0) \cos(\partial_x\theta_-)(0,y,\tau) \rangle=\frac{1}{(\tau^2+y^2)^{1/(K \pi^2)}}\nonumber\\
&\langle \cos(\partial_y\theta_-)(0,0,0) \cos(\partial_y\theta_-)(x,0,\tau) \rangle=\frac{1}{(\tau^2+x^2)^{1/(K \pi^2)}}
\end{align}
Notice that the dipole-$i$ correlation function is only nonzero when they are at the same row transverse to the dipole orientation. Thus, the dipoles effectively behave as a $1d$ Luttinger liquid with restricted motion and algebraic correlation on each stripe. This quasi-one-dimensional behavior is crucial for the existence of a critical point. As the dipole displays $1d$ motion within the same stripe, the quantum fluctuation forbids any dipole condensation with off-diagonal long-range order as a consequence of the Mermin-Wagner theorem so the critical point cannot be enlarged to a 2D superfluid phase. 

To explicate the relevance of boson compactification at QCP, we need to consider the instanton event which shifts the $\theta_-$ by $2N\pi$. In the dual representation, due to $\mathcal{T}$ symmetry which effectively sets the hardcore boson at half-filling, $\phi_-$ can not be uniform in space and are chosen to be integer and half-integers for different sites so we only consider the vertex operators $V=\cos( 4\pi \partial_i\phi_-),V'=\cos(2\pi\partial^2_i \phi_-)$ at IR limit. In Ref.~\cite{xu2008resonating,paramekanti2002ring}, it was shown that there is a finite region for $K>K_c$ where all vertex operators are irrelevant so the compactness of the boson can be ignored. 
As the Luttinger parameter is determined by the effective interaction, this requires the dipole to be weakly interacting and hence the $S_z S_z$ interaction should be relative small compared to the ring-exchange term and XY term. In our Hamiltonian in Eq.\ref{hhh}, we tune the $S_z S_z$ to zero so the
Luttinger parameter is large enough to escape from the instanton event. It is noteworthy to emphasize that the vertex operator $(-1)^y \cos(2\pi \partial_x \phi),(-1)^x\cos(2\pi \partial_y \phi)$ with staggered sign-factor would coarse grain to zero at long wave-length limit and hence can be ignored in the IR theory at the QCP. 

\section{Beyond the critical point}

The theory in Eq.~\ref{ac} can also be used to describe the gapped phases proximate to the QCP. When tuning away from the critical point, either the inter-site plaquette-ring exchange term or the intra-site XY term can trigger a gap via instanton proliferation\cite{gogolin1999bosonization} generated by the addition of 
\begin{align}
&(\lambda-\lambda_c)[\cos(2\pi \partial_x \phi)+\cos(2\pi \partial_y \phi)]
\label{defect1}
\end{align} to the dipole liquid theory.
Here $\phi$ is the vertex operator defined as $\partial_x\partial_y \phi=S^z=\hat{n}_b-1/2$ which triggers the $2\pi$ instanton tunneling of the compact boson.  Such a term is induced by the imbalance between intra-site an inter-site interactions, and is absent at the fine-tuned critical point. Once away from the critical region, the instanton proliferation gaps out the critical dipole liquid and yields either an onsite entangled state as the trivial Mott phase or an inter-site plaquette ordering as the HOTI phase.

When passing the critical point, the competition and imbalance between the onsite interaction versus the plaquette interaction triggers another vertex(instanton proliferation) term $(\lambda-\lambda_c)[ \cos(2\pi \partial_x \phi)+\cos(2\pi \partial_y \phi)]$.
Due to $\mathcal{T}$ symmetry which is akin to the hardcore boson at half-filling, $\phi$ cannot be chosen to be all integers on the dual sites and at least one of them among the four sites on the dual plaquette has to be a half-integer. In addition, the translation symmetry(here we consider the stretched lattice with each unit-cell enlarged to a plaquette as Fig.~\ref{app}) acts as,
\begin{align}
&T_x:\phi(x,y)\rightarrow \phi(x+1,y)+\frac{y}{2},~T_y:\phi(x,y)\rightarrow \phi(x,y+1)+\frac{x}{2},
\end{align}
This implies $\cos(2\pi \partial_x \phi)+\cos(2\pi \partial_y \phi)$ breaks lattice translation and inversion and thus drives imbalance between distinct plaquettes configurations. Such operator is absent at the QCP where the intra and inter unit cell coupling is balanced when fine-tune $\lambda=\lambda_c$.
However, once we move away from the critical point $\lambda_c$, the UV Hamiltonian generates such vertex terms and hence triggers to phase into a HOTI state with plaqutte order or a trivial Mott phase with on-site entangled patterns.

To visualize the relation between vertex operator and distinct plaquette patterns, we investigate the effect of vertex operator
$V_i=\cos( 2\pi \partial_i\phi_-)$. This operator creates a kink for the dipole-i along the transverse j-stripe and hence prompt a dipole gap which breaks translation $T_x,T_y$ but is still invariant under $\mathcal{T}$. 
From either symmetry argumens or from renormalization group flow of the lattice theory\cite{xu2007bond}, this term induces a plaquette order between two dipoles on the transverse stripe with four spins entangled at the corner of a square.
\begin{align} 
&\cos( 2\pi \partial_i\phi_-)\sim\cos( 2\pi \partial_y\partial_x \theta_-)e^{i\pi r_i},
\end{align}
Depending on the sign of $(\lambda-\lambda_c)$, the plaquette entangled ordering pattern either falls into the unit cell or the intra-site plaquette, which exactly matches the feature of trivial Mott or HOTI phase.

\end{document}